\documentclass[conference]{IEEEtran}
\IEEEoverridecommandlockouts
\usepackage{cite}
\usepackage{amsmath,amssymb,amsfonts}
\usepackage{algorithm}
\usepackage{algpseudocode}
\usepackage{graphicx}
\usepackage{psfrag}
\usepackage{subcaption}
\graphicspath{{./images/}}
\DeclareGraphicsExtensions{.pdf,.jpeg,.png}
\usepackage{textcomp}
\usepackage{xcolor}

\newcommand{\Hermitian}{\mathrm{H}}

\newcommand{\be}{\begin{equation}}
\newcommand{\ee}{\end{equation}}
\newcommand{\bse}{\begin{subequations}}
\newcommand{\ese}{\end{subequations}}
\newcommand{\bea}{\begin{eqnarray}}
\newcommand{\eea}{\end{eqnarray}}

\newcommand{\bF}{{\bf F}}

\newcounter{tempEquationCounter}
\newcounter{thisEquationNumber}
\makeatletter
\newcommand\fs@spaceruled{\def\@fs@cfont{\bfseries}\let\@fs@capt\floatc@ruled
  \def\@fs@pre{\vspace{0.5\baselineskip}\hrule height.8pt depth0pt \kern2pt}%
  \def\@fs@post{\kern2pt\hrule\relax}%
  \def\@fs@mid{\kern2pt\hrule\kern2pt}%
  \let\@fs@iftopcapt\iftrue}
\makeatother
\begin{document}

\title{Multiuser OTFS Channel Parameter Estimation Toward Grid-Independent Regime \\

\thanks{This publication has emanated from research supported in part by research grants from Research Ireland under the US-Ireland R\&D Partnership Programme Grant Numbers 24/US/4013 and 21/US/3757, and National Science Foundation under Grants NSF ECCS-2153875, CNS-2229562 and ECCS-2526413. For the purpose of Open Access, the authors have applied a CC BY public copyright license to any Author Accepted Manuscript version arising from this submission.
 }

}

\author{
\IEEEauthorblockN{Hanning Wang\IEEEauthorrefmark{1}, Rong-Rong Chen\IEEEauthorrefmark{2} and Arman Farhang\IEEEauthorrefmark{1}}

\IEEEauthorblockA{\IEEEauthorrefmark{1}Department of Electronic and Electrical Engineering, Trinity College Dublin, Ireland \\
\IEEEauthorrefmark{2}Department of Electrical and Computer Engineering, University of Utah, USA. \\} 

\IEEEauthorblockA{ Email: \{wangh15, arman.farhang\}@tcd.ie, rchen@utah.edu}}

\maketitle

\begin{abstract}
We study channel parameter estimation for multiuser orthogonal time frequency space (OTFS) systems in the delay-Doppler (DD) domain. To enable structured parametric estimation, we adopt a multi-user pilot cyclic prefix (MU-PCP) design, which multiplexes users along the Doppler dimension while preserving a separable exponential structure. This structure facilitates high-resolution estimation of fractional delay and Doppler parameters in the multiuser setting. Building on this framework, we extend weighted MUSIC (W-MUSIC) to multiuser OTFS, providing a computationally efficient approach with mild grid dependency, and develop a matrix pencil (MP)-based method that achieves fully grid-independent delay-Doppler parameter estimation. {Numerical results demonstrate the effectiveness of the proposed methods and reveal a robustness-complexity tradeoff: W-MUSIC performs better at low SNR, while MP achieves higher estimation accuracy at moderate-to-high SNR with significantly lower computational complexity.
}

\end{abstract}

\vspace{-0.4cm}
\section{Introduction}
Accurate channel estimation in time-varying wireless environments is critical for reliable communication, especially in high-mobility scenarios. In such settings, the channel is doubly selective in the time-frequency (TF) domain, limiting the effectiveness of conventional TF-domain methods. This motivates delay-Doppler (DD) domain signal processing, where orthogonal time frequency space (OTFS) provides a structured representation of the channel~\cite{OTFS}. In the DD domain, the channel can be modeled as a superposition of a small number of paths characterized by delay and Doppler shifts that evolve slowly over time~\cite{OTFSpredictability}. This naturally leads to a parametric representation, linking channel and parameter estimation which is key to integrated sensing and communication.

While OTFS has been extensively studied in single-user settings, practical wireless systems require multiuser support, where inter-user interference poses significant challenges for structured parameter estimation in the DD domain. To address this, we adopt a multi-user pilot with cyclic prefix (MU-PCP) which was originally proposed for synchronization in the uplink of multiuser OTFS \cite{mu_pcp}. MU-PCP multiplexes users' pilots along the Doppler dimension. This structure is well-suited for parametric estimation, as it enables the received signal to retain a form compatible with high-resolution techniques even in the presence of multiple users. Building on this property, in this paper, we extend parametric DD  estimation methods to multiuser uplink OTFS with MU-PCP. 

Existing DD domain channel estimation methods can be broadly classified into grid-based and grid-independent approaches. Grid-based methods discretize the DD domain and estimate the channel over a predefined dictionary, including threshold-based schemes \cite{embedpilotcs}, correlation-based estimators \cite{grid_correlation}, and compressed sensing (CS) techniques \cite{CSOMP1}. These methods rely on an on-grid assumption, causing basis mismatch and performance degradation with fractional delay and Doppler shifts \cite{interferencecancellation}. Techniques based on basis expansion model (BEM) \cite{bempilot} and dictionary refinement \cite{DA_OMP} mitigate leakage but do not directly estimate continuous delay and Doppler parameters.

Grid-independent approaches aim to directly estimate continuous DD parameters without relying on discretized grids. Subspace-based methods including root-MUSIC and ESPRIT exploit signal-noise subspace orthogonality but often incur high computational load 
\cite{root_music_otfs}. Weighted MUSIC (W-MUSIC) reduces this load by approximating polynomial coefficients and enables fractional DD estimation; however, it retains a mild form of grid dependency, as the polynomial roots are evaluated over discretized samples on the unit circle \cite{weighted_2d_music_delay_doppler}.
In contrast, the matrix pencil (MP) method fully exploits the exponential structure of the signal and directly estimates signal parameters via a generalized eigenvalue problem, achieving a truly grid-independent solution~\cite{mp_svd,2d_mp_toa_aoa}. The two methods exhibit complementary behavior across SNR regimes: MP demonstrates superior performance in moderate-to-high signal to noise ratio (SNR) regimes, while W-MUSIC exhibits higher robustness in low-SNR conditions. 
Despite their strong performance in single-user settings, extending these methods to multiuser OTFS remains non-trivial due to multiuser interference and lack of a suitable signal structure.

To address the above-mentioned research gaps, in this paper, we leverage the MU-PCP structure for grid-independent DD parameter estimation in multiuser OTFS. We extend W-MUSIC to the multiuser case, providing a computationally efficient approach with mild grid dependency. We develop a fully grid-independent MP-based estimator for delay and Doppler estimation. Our main contributions are as follows:

\begin{itemize}

    \item 
We introduce an MU-PCP-based pilot design and derive the resulting multiuser TF-domain signal model, which preserves a separable exponential structure with user-specific Doppler offsets. This structure enables user separation and reformulates multiuser OTFS channel estimation as a joint parametric problem over fractional delay and Doppler shifts, providing the foundation for grid-independent DD estimation.

     \item 
 We extend W-MUSIC to multiuser uplink by exploiting the structured model for subspace construction and polynomial approximation, with mild grid dependency.
  
     \item
We develop a fully grid-independent MP-based method using a structured block Hankel construction and reduced pencil formulation, enabling direct estimation of DD parameters with improved accuracy at moderate-to-high SNR and significantly lower complexity than W-MUSIC.

  \item 
  We compare MU-W-MUSIC and MU-MP through simulations with fractional delay and Doppler. The results reveal a robustness-complexity tradeoff: MU-W-MUSIC performs better at low SNR, while MU-MP achieves higher accuracy at moderate-to-high SNR with substantially reduced complexity.

\end{itemize}

The rest of the paper is structured as follows:  In Section~\ref{sec:multiuser otfs uplink} we present the multiuser OTFS uplink model. Section~\ref{sec:proposed_mu_estimation} introduces the proposed MU-W-MUSIC and MU-MP channel estimation methods. The complexity analysis and simulation results are discussed in Section~\ref{sec:results} and Section~\ref{sec:complexity}. Our conclusion is provided in Section~\ref{sec:conclusion}.

\section{Multiuser OTFS Uplink} \label{sec:multiuser otfs uplink}

In this section, we present the multiuser uplink OTFS system model with MU-PCP. We begin by reviewing the DD channel model, followed by the introduction of the MU-PCP structure for multiuser transmission. Based on this structure, we derive the corresponding transmitted and received signal models, which establish a structured foundation for the channel parameter estimation methods developed in Section \ref{sec:proposed_mu_estimation}.\footnote{Notations: In this paper, matrices, vectors and scalars are denoted as uppercase boldface, lowercase boldface letters and normal letters, respectively. $\mathbf{A}^{\rm T}$, $\mathbf{A}^{\Hermitian}$, $\mathbf{A}^{*}$ and $\mathbf{A}^{\dagger}$ represent the transpose, Hermitian transpose, conjugate and pseudo-inverse of the matrix, respectively. $a[i,j]$ denotes the element on $i$-th row, $j$-th column of the matrix $\mathbf{A}$. $\mathbf{I}_L$ denotes a $L \times L$ identity matrix. $\delta(\cdot)$ is the Dirac delta function. $g(\cdot)$ is the impulse response of the transmit pulse shape.
The function $\operatorname{diag}\{\mathbf{a}\}$ forms a diagonal matrix with the entries of vector $\mathbf{a}$ on the main diagonal. $\lfloor \cdot \rfloor$ and $\lceil \cdot \rceil$ represent the floor and ceiling calculation, respectively.}

\subsection{Delay-Doppler Channel Model} \label{sec:delay_doppler_channel_model}

We consider a discrete-time baseband transmit signal vector $\mathbf{s}=[s[0],\ldots,s[MN-1]]^{\rm T}\in \mathbb{C}^{MN \times 1}$ with $MN$ samples, spaced by $\Delta\tau$ seconds, where $M$ and $N$ denote the number of delay and Doppler bins on the DD plane. The total signal duration and bandwidth are given by $T_{\mathrm{total}} = MN\Delta\tau$ and $B = MN\Delta\nu$, where $\Delta\tau$ and $\Delta\nu = \frac{1}{MN\Delta\tau}$ represent the delay and Doppler resolutions. The DD domain multi-path channel response can be formulated as 
    \begin{equation}
        h(\tau, \nu)=\sum\nolimits_{i=0}^{P-1} h_i \delta\left(\tau-\tau_i\right) \delta\left(\nu-\nu_i\right), 
    \label{eq:multipathchannel_dirac}
    \end{equation}
where $P$ is the total number of propagation paths, and $h_i$, $\tau_i$ and $\nu_i$ are the complex channel gain, delay, and Doppler shift of the $i^{\rm th}$ path, respectively.  After sampling the channel at the Nyquist rate with sampling period $T_{\rm s} = \Delta\tau$, the normalized delay and Doppler shifts are denoted as $\ell_i={\tau_i}/{\Delta \tau}$, and $\kappa_i={\nu_i}/{\Delta \nu}$, where $\ell_i \in [0, \ell_{\max} - 1] $, $\kappa_i \in[-\kappa_{\max}/2, \kappa_{\max}/2] $,
and
$\ell_{\max}={\tau_{\max}}/{\Delta \tau}$ and $\kappa_{\max}={\nu_{\max}}/{\Delta \nu}$ are determined by maximum delay spread $\tau_{\max}$ and Doppler spread $\nu_{\max}$ of the channel. In this work, we consider  fractional delays $\ell_i$ and Dopplers $\kappa_i$. We add a cyclic-prefix (CP) with length $L_{\mathrm{cp}} \geq \ell_{\max}-1$ at the beginning of each block of $MN$ samples, i.e., $\mathbf{s}$, to avoid inter-block interference (IBI). After the transmit signal is passed through the channel and CP removal, the received signal is represented as
\begin{equation}
r[l] = \sum\nolimits_{i=0}^{P-1}  h_i \sum\nolimits_{l'=0}^{MN-1}\! s[l'] g\big((l - l' - \ell_i)\Delta \tau\big) e^{j \frac{2 \pi \kappa_i l}{MN}} + \eta[l],
\label{eq:received_signal_fractional}
\end{equation}
In this work, we consider $g(\cdot)$ an ideal sinc pulse-shape. Stacking the signal samples $r[l]$ in a vector $\mathbf{r}=[r[0],\ldots,r[MN-1]]^{\rm T}$, \eqref{eq:received_signal_fractional} can be represented in vectorized form as $\mathbf{r} = \mathbf{H} \mathbf{s} + \boldsymbol{\eta},$ where $\boldsymbol{\eta}\in\mathbb{C}^{MN \times 1}$ is the noise vector, with the elements following a circularly symmetric  complex Gaussian distribution, i.e., $\eta[n] \sim \mathcal{CN}(0,\sigma^2)$. The delay-time (DT) domain channel matrix, $\mathbf{H} \in \mathbb{C}^{MN \times MN}$, is constructed as ~\cite{channelmodel1,isachenk} $\mathbf{H}=\sum_{i=0}^{P-1} h_i \mathbf{\Pi}^{\ell_i}\mathbf{\Delta}^{\kappa_i}$, where $\mathbf{\Pi}^{\ell_i} \in \mathbb{C}^{MN \times MN}$ denotes the fractional delay operator matrix with a delay shift of $\ell_i$: $\mathbf{\Pi}^{\ell_i} = \mathbf{F}_{MN}^\Hermitian \mathbf{B}^{\ell_i}  \bF_{MN}$ where $\mathbf{F}_{MN} \in \mathbb{C}^{MN \times MN}$ is the $MN$-point normalized discrete Fourier transform (DFT) matrix, and $\mathbf{B}^{\ell_i} =\operatorname{diag}\left\{\mathbf{b}_M(\ell_i) \otimes \mathbf{b}_N(\ell_i)\right\} $ is a diagonal matrix with \cite{isachenk}
\begin{align}
& \mathbf{b}_M(\ell_i) \triangleq[1, e^{-j ({2 \pi}/{M}) \ell_i} ,\cdots , e^{-j ({2 \pi (M-1)}/{M}) \ell_i}]^{\rm{T}} \in \mathbb{C}^{M \times 1}, \nonumber \\
& \mathbf{b}_N(\ell_i) \triangleq[
1, e^{-j ({2 \pi}/{MN}) \ell_i} , \cdots , e^{-j ({2 \pi (N-1)}/{MN}) \ell_i}
]^{\rm{T}} \in \mathbb{C}^{N \times 1}. \nonumber
\end{align}
The Doppler shift matrix $\boldsymbol \Delta ^{\kappa_i}  \in \mathbb{C}^{MN \times MN}$ is structured as  $\boldsymbol{\Delta}^{\kappa_i} =\operatorname{diag}\left\{\mathbf{v}_N(\kappa_i) \otimes \mathbf{v}_M(\kappa_i)\right\},$
where
\begin{align}
& \mathbf{v}_N(\kappa_i) \triangleq[1, e^{j ({2 \pi}/{N})  \kappa_i} ,\cdots , e^{j ({2 \pi (N-1)}/{N}) \kappa_i}]^{\rm{T}} \in \mathbb{C}^{N \times 1}, \nonumber \\
& \mathbf{v}_M(\kappa_i) \triangleq[
1, e^{j ({2 \pi}/{MN}) \kappa_i} , \cdots , e^{j ({2\pi\!(M-1)}/{MN}) \kappa_i}
]^{\rm{T}} \in \mathbb{C}^{M \times 1}. \nonumber
\end{align}

\subsection{MU-PCP Structure and Transmitted Signal}\label{sec:MU-PCP Pilot Structure and Transmitted Signal}

At the transmitter, we adopt the MU-PCP structure introduced in \cite{mu_pcp}. This structure extends the pilot design in \cite{bempilot, weighted_2d_music_delay_doppler} to the multiuser setting. While \cite{mu_pcp} applies it to multiuser synchronization, in this work we employ the MU-PCP structure for multiuser OTFS channel parameter estimation.

To prevent Doppler-domain inter-user interference under the MU-PCP structure, the number of supported users must satisfy $Q \le \lfloor {N}/(2\kappa_{\max}+1) \rfloor$, where $\kappa_{\max}$ is the maximum Doppler shift \cite{mu_pcp}. For each user $q = 0, 1, \ldots, Q-1$, a pilot sequence is constructed using a Zadoff-Chu (ZC) sequence $x_{\mathrm{ZC}}$ of length $M_{\mathrm{ZC}}$, where the last $M_\mathrm{CP}$ elements are copied to the beginning of the sequence as a CP, resulting in a total length $M_{\mathrm{PCP}} = M_{\mathrm{CP}} + M_{\mathrm{ZC}}$, where $M_{\mathrm{PCP}} < M$. The CP length $M_{\mathrm{CP}} \ge \lceil \ell_{\max} \rceil$ . All users share the same delay support and may reuse the same ZC sequence, while being separated along the Doppler dimension. The Doppler index for user $q$ is given by $k^{q} = \frac{1}{2}\left\lfloor \frac{N}{Q} \right\rfloor + q \left\lfloor \frac{N}{Q} \right\rfloor$.
The DD domain pilot pattern for a given user $q$ is
\begin{equation}
x^{q}[l,k] =
\begin{cases}
x_{\mathrm{CP}}[l], & 0 \!\le l \le \!M_{\mathrm{CP}}-1,\ k = k^{q}, \\
x_{\mathrm{ZC}}[l], & M_{\mathrm{CP}}\! \le l \!\le\! M_{\mathrm{PCP}}\!-\!1,\ k = k^{q}, \\
0, & \text{elsewhere}.
\end{cases}
\label{eq:pcp_structure}
\end{equation}
We assume that sufficient guard intervals are inserted in the DD grid such that data induced interference can be neglected in the pilot region.

Following the Zak transform-based OTFS formulation in ~\cite{weighted_2d_music_delay_doppler}, the transmitted pilot for user $q$ can be written as
\begin{align}   
    s^{q}(t) = & \frac{1}{\sqrt{N}}\!\sum_{l=0}^{M-1}\!\sum_{k = 0}^{N-1} \! x^q[l,k]\!\!\sum_{n= 0}^{N-1} g(t-\!l\Delta\tau\!-nT)e^{j\frac{2 \pi kn}{N}} \notag \\
    = & \frac{1}{\sqrt{N}} \!\! \sum_{n = 0}^{N-1}\!\sum_{l=0}^{M_{\rm PCP}-1} \!  \!\!\! x^q[l,k^q]g(t-\!l\Delta\tau\!-nT)e^{j\frac{2 \pi k^{q}n}{N}}, \notag \\
    = & \sum_{n = 0}^{N-1}\widetilde{x}^q(t-nT) e^{j\frac{2 \pi k^{q}n}{N}}     \label{eq:transmitted_pilot_time}
\end{align}
where $T = M\Delta\tau$ is the Zak transform step length, and $\widetilde{x}^q(t) \triangleq \frac{1}{\sqrt{N}} \sum_{l=0}^{M_\mathrm{PCP}-1} x^q[l, k^q]g(t-l \Delta \tau)$.

\subsection{Received Signal Model}\label{sec:Received Signal Model}

We assume that the number of propagation paths for user $q$ is $P_q$. The uplink received signal can be expressed as
\begin{equation}
    r(t) = \sum\nolimits_{q=0}^{Q-1}\sum\nolimits_{i_q=0}^{P_q -1} h_{i_q}s^{q}(t - \tau_{i_q})e^{j2\pi \nu_{i_q} t} + \eta(t), 
    \label{eq:rx_continous_time} 
\end{equation}
where $h_{i_q}$, $\tau_{i_q}$, and $\nu_{i_q}$ denote the complex path gain, delay, and Doppler shift of the $i_q$-th path associated with user $q$, respectively. Substitute \eqref{eq:rx_continous_time} into \eqref{eq:received_signal_fractional}, we obtain
\begin{equation}
r[l] \approx \sum\nolimits_{q=0}^{Q-1} \sum\nolimits_{i_q=0}^{P_q-1} 
h_{i_q}s^q\left((l-\ell_{i_q})\Delta\tau \right)
e^{j  \frac{2\pi\kappa_{i_q} \!n}{N}} + \eta[l].
\label{eq:rx_approx}
\end{equation}

The MU-PCP structure in \eqref{eq:transmitted_pilot_time} introduces a user-specific Doppler shift $k_{\mathrm p}^{q}$ in the transmitted signal. Substituting \eqref{eq:transmitted_pilot_time} into \eqref{eq:rx_approx}, the Doppler exponential term becomes $e^{j{2\pi \kappa_{i_q} n}/{N}} \cdot e^{j{2\pi k^{q} n}/{N}}
= e^{j{2\pi (\kappa_{i_q} + k^{q}) n}/{N}}$,
which shows that each user's paths are shifted to distinct Doppler regions. The approximation in \eqref{eq:rx_approx} is based on the fact that pilot length $M_\mathrm{PCP} \cdot \Delta \tau$ is much smaller than the duration $T = M\Delta \tau$, the Doppler-induced phase term $e^{j2\pi \nu_{i_q} t}$ can be approximated as constant over $t \in [nT, nT + M_\mathrm{PCP}\Delta \tau], n = 0, \ldots, N-1$ ~\cite{weighted_2d_music_delay_doppler, bempilot}.

After CP removal, the received pilot samples in the DT domain are arranged into a matrix $\mathbf{R}_{\mathrm{DT}} \in \mathbb{C}^{M_{\mathrm{ZC}} \times N}$. An $M$-point DFT is then applied along the delay dimension to obtain the time-frequency (TF) domain observation $\mathbf{R}_{\mathrm{TF}} \in \mathbb{C}^{M \times N}$. {By incorporating the MU-PCP structure into the received signal model and extending the TF-domain formulation in \cite{weighted_2d_music_delay_doppler}, we then express the TF-domain signal as}
\begin{equation}
\mathbf{R}_{\mathrm{TF}}
=
\sum\nolimits_{q=0}^{Q-1}\sum\nolimits_{i_q=0}^{P_q-1}
h_{i_q}\,
\widetilde{\mathbf{X}}_{\mathrm{f}} \mathbf{b}_M(\ell_{i_q})
\mathbf{v}_N^{\mathrm T}(\kappa_{i_q} + k^{q})
+
\mathbf{W},
\label{eq:effective_RTF_model}
\end{equation}
where $\widetilde{\mathbf{X}}_{\mathrm{f}} = \operatorname{diag}(\mathbf{x}_{\mathrm{f}})$ with $\mathbf{x}_{\mathrm{f}} \in \mathbb{C}^{M \times 1}$ denoting the $M$-point DFT of the ZC sequence $x_{\mathrm{ZC}}[l]$ in \eqref{eq:pcp_structure}, $\mathbf{W}$ is the noise matrix, and $\mathbf{b}_M(\ell_{i_q}) \in \mathbb{C}^{M \times 1}$ and $\mathbf{v}_N(\kappa_{i_q}) \in \mathbb{C}^{N \times 1}$ are the delay and Doppler steering vectors defined in Section \ref{sec:delay_doppler_channel_model}, respectively. 

Eq.~\eqref{eq:effective_RTF_model} shows that, under the MU-PCP structure, the received pilot in the TF domain admits a separable complex exponential form, where delay and Doppler shifts appear as separable exponentials along the frequency and time dimensions, while user separability is maintained through structured Doppler offsets. In the following sections, we estimate the parameters $h_{i_q}$, $\ell_{i_q}$, and $\kappa_{i_q}$ based on this model.

\section{Proposed Multiuser Channel Estimation} \label{sec:proposed_mu_estimation}

In this section, we develop two channel parameter estimation methods based on the structured signal model in \eqref{eq:effective_RTF_model}, by extending W-MUSIC to the multiuser setup (MU-W-MUSIC). We propose MU-MP  that  exploits the underlying exponential structure. Although W-MUSIC and MP are established high-resolution parameter estimation tools, their direct application to multiuser OTFS signal model is not immediate. Under the MU-PCP structure, each physical Doppler $\kappa_{i_q}$ is observed as $\kappa_{i_q} + k_q$ with user-specific offset $k_q$. Consequently, our  methods jointly estimate all $P_{\mathrm{tot}}$ paths, associate the observed Dopplers with users via the MU-PCP partitions, remove the offsets, and recover channel gains using a multiuser parametric dictionary. The MU-W-MUSIC and MU-MP methods are detailed in Sections III-A and III-B, respectively, followed by channel gain estimation in Section III-C.

\vspace{-0.2cm}
\subsection{Proposed MU-W-MUSIC Based Parameter Estimation} \label{sec:mu_w_music}

In the received signal model \eqref{eq:effective_RTF_model}, user separation is embedded through the Doppler offsets $k_p^q$. This structured representation enables the extension of the single user W-MUSIC methods to the multiuser setting.

\subsubsection{Subspace Construction}

Following \cite{weighted_2d_music_delay_doppler}, spatial smoothing is applied to the TF-domain received signal to generate $L_{\mathrm{snap}} = N - N' + 1$ snapshots. The $j$-th snapshot is obtained by extracting an $M' \times N'$ submatrix of $\mathbf{R}_{\mathrm{TF}}$ along the Doppler dimension and vectorizing it as $\mathbf{r}_j = \operatorname{vec}(\mathbf{R}_{\mathrm{TF}}^{(j)})$. The sample covariance matrix is then computed as $\widehat{\mathbf{K}} = ({1}/{L_{\mathrm{snap}}}) \sum_{j=1}^{L_{\mathrm{snap}}} \mathbf{r}_j \mathbf{r}_j^{\Hermitian}$. Followed by eigenvalue decomposition (EVD)  The noise subspace $\mathbf{E}_n$ is formed by excluding the $P_{\mathrm{tot}}$ dominant eigenvectors of $\widehat{\mathbf{K}}$, where $P_{\mathrm{tot}} = \sum_{q=0}^{Q-1} P_q$ denotes the total number of propagation paths among all users.

\subsubsection{Doppler Estimation}

We define the effective noise space projection matrix as
\begin{equation}
\mathbf{E}_{\mathrm{eff}} = (\mathbf{I}_{N'} \otimes \widetilde{\mathbf{X}}_{\mathrm{f}})^H
\mathbf{E}_n\mathbf{E}_n^H
(\mathbf{I}_{N'} \otimes \widetilde{\mathbf{X}}_{\mathrm{f}}).
\label{eq:effective_noise_sub}
\end{equation}
$\mathbf{v}_{N'}(z_\kappa)$ is obtained by substitute $z_\kappa = e^{j \frac{2\pi}{N}\kappa}$ into $\mathbf{v}_{N}(\kappa)$ defined in section \ref{sec:delay_doppler_channel_model}, and retain its first $N'$ entries. The Doppler null-spectrum projection matrix is constructed as
\begin{equation}
\mathbf{D}(z_\kappa) = \! \left(\mathbf{v}_{N'}(z_\kappa)\otimes \mathbf{I}_{M'}\right)^\Hermitian
\mathbf{E}_{\mathrm{eff}}
\left(\mathbf{v}_{N'}(z_\kappa)\otimes \mathbf{I}_{M'}\right) ,
\label{eq:Doppler_null_spec}
\end{equation}
where the roots of the polynomial $D(z_\kappa) = \operatorname{det}[\mathbf{D}(z_\kappa)] =0$ correspond to the Doppler shifts of all propagation paths.

The W-MUSIC approximates the polynomial coefficients via a truncated Fourier series \cite{og_w_r_music,weighted_2d_music_delay_doppler}. This is achieved by evaluating $D(z_\kappa)$ at $z_{\kappa,m}=e^{j\phi_m}$ with $\phi_m \in [-\pi,\pi]$ along the unit circle, $m=0,\ldots,Q_{\rm sample}-1$. Note that this sampling is used solely for polynomial approximation and does not impose a discretization on the Doppler parameter space.

Let $\mathbf{d} = [D(z_{\kappa,0}), \ldots, D(z_{\kappa,Q_{\rm sample}-1})]^{\rm T}$, The approximation $D(e^{j\phi_m}) \approx \sum_{g=-G}^{G} f_g e^{j\phi_m g}$
can be written in matrix form as $\mathbf{d} \approx \mathbf{\Phi}\mathbf{f}$, where $\mathbf{\Phi} \in \mathbb{C}^{Q_{\rm sample} \times (2G+1)}$ with entries $[\mathbf{\Phi}]_{m,g} = e^{j\phi_m (g-G)}$. The coefficient vector $\mathbf{f}$ is then obtained via a weighted least squares (WLS) solution:
\begin{equation}
\widehat{\mathbf{f}} = (\mathbf{\Phi}^{\Hermitian}\mathbf{\Gamma}\mathbf{\Phi})^{-1}
\mathbf{\Phi}^{\Hermitian}\mathbf{\Gamma}\mathbf{d},
\label{eq:wls_coefficient_music}
\end{equation}
where $\boldsymbol{\Gamma}=\operatorname{diag}(\gamma_0,\ldots,\gamma_{Q_{\rm sample}-1})$ is a weighting matrix with $\gamma_m = \frac{1}{|[\mathbf{d}]_m|+\epsilon}$. This yields a low-order approximation of the Doppler polynomial $\widehat{D}(z_\kappa)$.

Next, the $P_{\mathrm{tot}} = \sum_{q=0}^{Q-1} P_q$ roots $\widehat{z}_{\kappa,i}$ of $\widehat{D}(z_\kappa)$ that lie inside and are closest to the unit circle are selected. The corresponding Doppler estimates are then computed as 
\begin{equation}
\widehat{\kappa}_{i}^\mathrm{(obs)} = ({N}/{2\pi})\angle(\widehat{z}_{\kappa,i}).
\end{equation}
Here, we denote $\widehat{\kappa}_i^{\mathrm{(obs)}} = \widehat{\kappa}_{iq} + k_q$ as the observed Doppler shifts, which contains both the physical Doppler $\kappa_{iq}$ and the user-specific Doppler offset $k_q$ introduced by the MU-PCP structure.
The estimated Doppler shifts are then grouped according to the index $k_q$ with each user, i.e., $\mathcal{K}_q = \left\{ \widehat{\kappa}_{i_q}^{\mathrm{(obs)}} \;\mid \; \widehat{\kappa}_i^{\mathrm{(obs)}} \in [k^{q} - \kappa_{\max},\, k^{q} + \kappa_{\max}] \right\},$ and the per-user Doppler parameters are calculated as
\begin{equation}
\widehat{\kappa}_{i_q} = \widehat{\kappa}_{i_q}^{\mathrm{(obs)}}- k^{q}, \quad \widehat{\kappa}_{i} \in \mathcal{K}_q.
\label{eq:user_Doppler_set}
\end{equation}

\subsubsection{Delay Estimation}

The delay polynomial corresponding to each Doppler root $\{\widehat{z}_{\kappa_{i_q}}\}_{i_ q = 0}^{P_q}$ is constructed as
\begin{equation}
J(z_\ell)
=
\mathbf{b}_{M'}^{\Hermitian}(z_\ell)\,
\mathbf{D}(\widehat{z}_{\kappa_{i_q}})\,
\mathbf{b}_{M'}(z_\ell),
\label{eq:delay_polynomial}
\end{equation}
where $\mathbf{b}_{M'}(z_\ell)$ is obtained by substituting $z_\ell = e^{j \frac{2\pi}{M}\ell}$ into $\mathbf{b}_{M}(\ell)$ defined in Section \ref{sec:delay_doppler_channel_model}, and retaining its first $M'$ entries. 

For each Doppler root $\widehat{z}_{\kappa_{i_q}}$ , the delay roots $\widehat{z}_{\ell,i_q}$ is selected as the roots of $J(z_\ell)$ that lie inside and closest to the unit circle. The corresponding delay estimates are then obtained as
\begin{equation}
\widehat{\ell}_{i_q}
=
-({M}/{2\pi})\angle(\widehat{z}_{\ell,i_q}).
\label{eq:delay_root_per_user}
\end{equation}

\subsection{Proposed MU-MP Based Parameter Estimation} \label{sec:mu_mp}

Building upon the structured signal model in \eqref{eq:effective_RTF_model}, in this section we propose a multiuser Matrix Pencil (MU-MP) based parameter estimation method.

\subsubsection{Block Hankel Matrix Construction}

We first construct a Doppler dimension Hankel matrix by arrange the $n$-th column of $\mathbf{R}_\mathrm{TF} \in \mathbb{C}^{M \times N}$ as
\begin{equation}
\mathbf{R}_n =
\begin{bmatrix}
r[0,n] & r[1,n] & \cdots & r[K_M-1,n] \\
r[1,n] & r[2,n] & \cdots & r[K_M,n] \\
\vdots & \vdots & \ddots & \vdots \\
r[M_{\mathrm p}-1,n] & r[M_{\mathrm p},n] & \cdots & r[M-1,n]
\end{bmatrix}
,
\label{eq:doppler_direction_inner_hankel}
\end{equation}
where $\mathbf{R}_n \in \mathbb{C}^{M_{\mathrm p}\times K_M}$, $M_{\mathrm p}$ and $N_{\mathrm p}$ are the pencil parameters along the delay and Doppler dimensions, respectively, with $K_M = M - M_{\mathrm p} + 1, K_N = N - N_{\mathrm p} + 1$. These matrices are then arranged into a block Hankel structure as:
\begin{equation}
\mathbf{X} =
\begin{bmatrix}
\mathbf{R}_0 & \mathbf{R}_1 & \cdots & \mathbf{R}_{K_N-1} \\
\mathbf{R}_1 & \mathbf{R}_2 & \cdots & \mathbf{R}_{K_N} \\
\vdots & \vdots & \ddots & \vdots \\
\mathbf{R}_{N_{\mathrm p}-1} & \mathbf{R}_{N_{\mathrm p}} & \cdots & \mathbf{R}_{N-1}
\end{bmatrix}
.
\label{eq:doppler_direction_block_hankel}
\end{equation}

\subsubsection{Doppler Estimation}

In the noiseless case, \eqref{eq:doppler_direction_block_hankel}  could be written as \cite{mp_svd,2d_mp_toa_aoa}: $\mathbf{X}
=
\mathbf{L}_{\nu}\boldsymbol{\Xi}\mathbf{R}_{\nu}^{\mathrm T}$, where $\boldsymbol{\Xi}=\operatorname{diag}\{h_0,\ldots,h_{P_{\mathrm{tot}}-1}\}$, and
\begin{equation*}
\mathbf{L}_{\nu}
=
\big[
\mathbf{B}_{M_{\mathrm p}},\;
\mathbf{B}_{M_{\mathrm p}}\mathbf{D}_{\nu},\;
\cdots,\;
\mathbf{B}_{M_{\mathrm p}}\mathbf{D}_{\nu}^{N_{\mathrm p}-1}
\big]^{{\mathrm T}}
\in \mathbb{C}^{M_{\mathrm p}N_{\mathrm p}\times P_{\mathrm{tot}}},
\end{equation*}
\vspace*{-0.5cm}
\begin{equation*}
\mathbf{R}_{\nu}
=
\big[
\mathbf{B}_{K_M},\;
\mathbf{B}_{K_M}\mathbf{D}_{\nu},\;
\cdots,\;
\mathbf{B}_{K_M}\mathbf{D}_{\nu}^{K_N-1}
\big]^{{\mathrm T}}
\in \mathbb{C}^{K_MK_N\times P_{\mathrm{tot}}},
\end{equation*}
where $\mathbf{B}_{M_{\mathrm p}}$ and $\mathbf{B}_{K_M}$ are obtained by retaining the first $M_{\mathrm p}$ and $K_M$ rows of $\mathbf{B} = \big[\mathbf{b}_M(\ell_0),\ldots,\mathbf{b}_M(\ell_{P_{\mathrm{tot}}-1})\big] \in \mathbb{C}^{M \times P_\mathrm{tot}}$, respectively. The Doppler pole matrix is $\mathbf{D}_{\nu}= \operatorname{diag}
\left\{
z_{\nu,0},z_{\nu,1},\ldots,z_{\nu,P_{\mathrm{tot}}-1}
\right\},$
with $z_{\nu,i}
=
e^{j\frac{2\pi}{N}(\kappa_i+k_p^{q_i})}$.

Partition the Hankel matrix $\mathbf{X}$ into $K_N$ block columns as
\begin{equation}
\mathbf{X}
=
\begin{bmatrix}
\mathbf{X}^{(0)} & \mathbf{X}^{(1)} & \cdots & \mathbf{X}^{(K_N-1)}
\end{bmatrix},
\label{eq:block_partition}
\end{equation}
where each block $\mathbf{X}^{(j)} \in \mathbb{C}^{M_{\mathrm p}N_{\mathrm p}\times K_M}$, $j=0,\ldots,K_N-1$. Next, the left and right pencil matrices, $\mathbf{X}_{\mathrm l}$ and 
$\mathbf{X}_{\mathrm r}$, are formed by removing the last and first block in \eqref{eq:block_partition}, i.e. $\mathbf{X}^{(K_N-1)}$ and $\mathbf{X}^{(0)}$, respectively. Due to the Doppler-domain shift-invariance, the two matrices satisfy
\begin{equation}
\mathbf{X}_{\mathrm l} = \mathbf{L}_{\nu}\boldsymbol{\Xi}\mathbf{R}_{\nu,o}^{\mathrm T}, \ \ 
\mathbf{X}_{\mathrm r} = \mathbf{L}_{\nu}\boldsymbol{\Xi}\mathbf{D}_{\nu}\mathbf{R}_{\nu,o}^{\mathrm T},
\end{equation}
where the truncated matrix $\mathbf{R}_{\nu,o}$ is obtained by removing the last $K_N$ rows of $\mathbf{R}_\nu$. The matrix pencil is formed as
\begin{equation}
\mathbf{X}_{\mathrm r}-\lambda\mathbf{X}_{\mathrm l}
=
\mathbf{L}_{\nu}\boldsymbol{\Xi}
\left(\mathbf{D}_{\nu}-\lambda\mathbf{I}_{P_{\mathrm{tot}}}\right)
\mathbf{R}_{\nu,o}^{\mathrm T}.
\label{eq:matrix_pencil}
\end{equation}

Given that $\mathbf{L}_{\nu}$ and $\mathbf{R}_{\nu,o}$ have full column rank, the rank of
$\mathbf{X}_{\mathrm r}-\lambda\mathbf{X}_{\mathrm l}$ drops to $P_\mathrm{tot} - 1$ when
$\lambda_i=z_{\nu,i}$ for $i = 0, 1, \ldots, P_\mathrm{tot} - 1$. Hence, the Doppler poles are obtained as the generalized eigenvalues of the pencil pair
$(\mathbf{X}_{\mathrm r},\mathbf{X}_{\mathrm l})$.

As it shown in \cite{mp_svd}, we can find the generalized eigenvalues of $(\mathbf{X}_{\mathrm r},\mathbf{X}_{\mathrm l})$ by first performing singular value decomposition (SVD) on $\mathbf{X}_{\mathrm l} = \mathbf{U}_{s}\boldsymbol{\Sigma}_{s}\mathbf{V}_{s}^{\mathrm H} + \mathbf{U}_{n}\boldsymbol{\Sigma}_{n}\mathbf{V}_{n}^{\mathrm H}$, where $\boldsymbol{\Sigma}_{s}$, $\mathbf{U}_{s}$ and $\mathbf{V}_{s}$ contain the largest $P_{\mathrm{tot}}$ singular values and their corresponding left and right singularvectors. Then we obtain the reduced pencil matrix as
\begin{equation}
\mathbf{T}_{\nu}
=
\boldsymbol{\Sigma}_{s}^{-1}
\mathbf{U}_{s}^{\mathrm H}
\mathbf{X}_{\mathrm r}
\mathbf{V}_{s}, 
\end{equation}
where the eigenvalues of $\mathbf{T}_{\nu}$ provide the Doppler pole estimates $\{\lambda_i = \hat{z}_{\nu,i}\}_{i=0}^{P_{\mathrm{tot}}-1}$ as
\begin{equation}
\widehat{\kappa_i}^\mathrm{(obs)}
=
({N}/{2\pi})\angle(\hat{z}_{\nu,i}).
\end{equation}
The per-user Doppler parameters $\widehat{\kappa}_{i_q}$ are then recovered following the process \eqref{eq:user_Doppler_set}.

\subsubsection{Delay Estimation}

Define $\widehat{\mathbf{\Theta}}_N =
\left[
\mathbf{v}_N^\mathrm{(obs)}(\widehat{\kappa}_0), \ldots, \mathbf{v}_N^\mathrm{(obs)}(\widehat{\kappa}_{P_{\mathrm{tot}}-1})
\right].$ Based on \eqref{eq:effective_RTF_model}, the received signal can be written as
\begin{equation}
\mathbf{R}_{\mathrm{TF}} = \mathbf{G}\widehat{\mathbf{\Theta}}_N^{\mathrm T} + \mathbf{W},
\label{eq:mp_subspace}
\end{equation}
where $\widehat{\mathbf{\Theta}}_N$ spans the Doppler subspace, and $\mathbf{G} \in \mathbb{C}^{M \times P_{\mathrm{tot}}}$ contains the corresponding delay subspace components.

The matrix $\mathbf{G}$ is obtained by projecting $\mathbf{R}_{\mathrm{TF}}$ onto the Doppler subspace via least squares
\begin{equation}
\widehat{\mathbf{G}} =
\mathbf{R}_{\mathrm{TF}} \widehat{\mathbf{\Theta}}_N^{*}
\left(\widehat{\mathbf{\Theta}}_N^{\mathrm T}\widehat{\mathbf{\Theta}}_N^{*}\right)^{-1}.
\label{eq:mp_g_matrix}
\end{equation}
Each column of $\widehat{\mathbf{G}}$ lies in the delay subspace and the phase difference across adjacent rows is constant. Therefore, the delay parameters can be estimated as
\begin{equation}
\widehat{\ell}^{\mathrm{(obs)}}_{i} =
-\frac{M}{2\pi}
\angle\!\left(
\frac{\widehat{G}[m+1,i]}{\widehat{G}[m,i]}
\right), \quad \forall\, m \in [0, M-1].
\label{eq:mp_delay_est}
\end{equation}
Then the delay estimates are assigned to each user according to the Doppler partitioning in \eqref{eq:user_Doppler_set}.

\subsection{Channel Gain Estimation}

After estimating the DD parameters $\{\widehat{\ell}_{i_q}, \widehat{\kappa}_{i_q} \}$ for all users based on the proposed methods from \ref{sec:mu_w_music} and \ref{sec:mu_mp}, the channel gains are recovered via a least squares (LS) formulation based on the signal model in \eqref{eq:effective_RTF_model}.

\subsubsection{Dictionary Construction}

Using the estimated parameters, we construct a parametric dictionary that captures the contribution of each propagation path. For each path $i_q$, the corresponding atom is defined as
\begin{equation}
\mathbf{a}_{i_q}
=
\operatorname{vec}\!\left(
\widetilde{\mathbf{X}}_{\mathrm{f}} 
\mathbf{b}_M(\widehat{\ell}_{i_q})
\mathbf{v}_N^{\mathrm T}(\widehat{\kappa}_{i_q} + k^q)
\right).
\end{equation}
Stacking all $P_{\mathrm{tot}}$ atoms, we obtain the dictionary matrix
\begin{equation}
\mathbf{\Psi} =
\left[
\mathbf{a}_{0},\,
\mathbf{a}_1,\,
\ldots,\,
\mathbf{a}_{P_{\mathrm{tot}}-1}
\right]
\in \mathbb{C}^{MN \times P_{\mathrm{tot}}}.
\end{equation}

\subsubsection{Least Squares Gain Estimation}

Vectorizing the received signal as $\mathbf{r}_\mathrm{TF} = \operatorname{vec}(\mathbf{R}_{\mathrm{TF}})$,  \eqref{eq:effective_RTF_model} can be re-written as $\mathbf{r}_\mathrm{TF} = \mathbf{\Psi}\mathbf{h} + \mathbf{w},$
where $\mathbf{h} \in \mathbb{C}^{P_{\mathrm{tot}} \times 1}$ contains the channel gains of all paths. The channel gains are then obtained via least squares
\begin{equation}
\widehat{\mathbf{h}} = (\mathbf{\Psi}^{\Hermitian}\mathbf{\Psi})^{-1} \mathbf{\Psi}^{\Hermitian}\mathbf{r}_\mathrm{TF}.
\end{equation}

\section{Complexity Analysis}\label{sec:complexity}

In this section, we compare the computational cost of  MU-W-MUSIC and MU-MP. We measure computational complexity in terms of the number of complex multiplications (CMs), and denote by $C_{(\cdot)}$ the corresponding complexity of each processing stage. The analysis focuses on the dominant operations that determine the overall computational cost.

\subsection{MU-W-MUSIC Complexity}

The steps inducing dominant CMs in MU-W-MUSIC are
\begin{itemize}
\item \textit{Covariance matrix EVD:} $C_{\mathrm{EVD}}\!\sim\!\mathcal{O}((M'N')^3)$.

\item \textit{Doppler estimation:} $C_{\mathrm{spec}}\!\sim\!\mathcal{O}(Q_{\mathrm{sample}} M'(M'N')^2)$. 

\item \textit{Delay estimation:} $C_{\mathrm{delay}}\!\sim\!\mathcal{O}(P_{\mathrm{tot}} M'(M'N')^2)$.
\end{itemize}
Thus $C_{\mathrm{WM}}\!\sim\!\mathcal{O}((M'N')^3 + (Q_{\mathrm{sample}}+P_{\mathrm{tot}})M'(M'N')^2)$, increasing with both $Q_{\mathrm{sample}}$ and $P_{\mathrm{tot}}$.

\subsection{MU-MP Complexity}

The complexity of MU-MP is dominated by:

\begin{itemize}
\item \textit{SVD:} $C_{\mathrm{SVD}}\!\sim\!\mathcal{O}(mn^2+n^3)$.

\item \textit{Doppler estimation:} $C_{\mathbf{T}}\!\sim\!\mathcal{O}(P_{\mathrm{tot}}^2K_MK_N)$

\item \textit{Delay estimation:} $C_{\mathrm{proj}}\!\sim\!\mathcal{O}(MN P_{\mathrm{tot}} + N P_{\mathrm{tot}}^2 + P_{\mathrm{tot}}^3)$. 
\end{itemize}
Since $P_\mathrm{tot} \ll M_\mathrm{p} N_\mathrm{p}$, the dominant part $C_{\mathrm{MP}}\!\sim\!\mathcal{O}(M_{\mathrm{p}}N_{\mathrm{p}}(K_MK_N)^2 + (K_MK_N)^3)$. The dependence on $P_{\mathrm{tot}}$ appears only in lower-order terms.

\subsection{Complexity Comparison} \label{sec:complexity comparison}
Assuming $P_\mathrm{tot} \ll MN$ with typical parameter choices $M_{\mathrm p} \approx 0.9M$, $N_{\mathrm p} \approx 0.125N$, $M' \approx 0.5M$, and $N' \approx 0.3N$ ~\cite{og_w_r_music, mp1995og}, 
the number of CMs can be expressed as $C_{\mathrm{WM}}(M,N) \approx 0.025 (MN)^3 + 0.135 M^3 N^2$ and $C_{\mathrm{MP}}(M,N) \approx 0.003 (MN)^3$. {As shown in Table \ref{tab:complexity_comparison}}, this comparison shows that MU-MP requires approximately 10 times fewer CMs than MU-W-MUSIC.

\section{Simulation Results}\label{sec:results}
We consider a multiuser OTFS system with $M=32$, $N=64$, and $Q=4$ users, assuming perfect synchronization and power control. Each user employs a pilot of length $M_{\mathrm{PCP}}=12$ with $L_{\mathrm{cp}}=4$ and $M_{\mathrm{ZC}}=8$. The number of paths per user $P_q$ is randomly selected between 1 and 3, with channel gains $|h_{i_q}|$ take random values between 0 and 1. Both integer and fractional delay and Doppler are considered, where path delays $\ell_{i_q} $ are uniformly distributed within the range $[0, \ell_{\max} - 1]$, while the Doppler shifts $\kappa_{i_q}$ are uniformly generated in the range $[-\kappa_{\max}/2, \kappa_{\max}/2]$, with $\ell_{\max}=4$ and $\kappa_{\max}=6$. For the MU-MP method, we choose $M_\mathrm{p} = 30$, $N_\mathrm{p} = 16$. For the MU-W-MUSIC approach, we select $M' = 16$, $N' = 20$, $G = 51$, and $Q_\mathrm{sample} = 128$. The root mean square error (RMSE) for each parameter $\theta \in \{\ell,\kappa,h\}$ is computed as $\mathrm{RMSE}_\theta = \sqrt{({1}/{P})\sum\nolimits_{i=0}^{P-1} (\theta_i - \hat{\theta}_i)^2}.$ For the channel matrix, the RMSE is calculated using the Frobenius norm as
$\mathrm{RMSE}_{\mathbf{H}} =
\sqrt{({1}/{MN}) \| \mathbf{H} - \widehat{\mathbf{H}} \|_F^2},$
where $\mathbf{H}$ and $\widehat{\mathbf{H}}$ denote the true and estimated channel matrices, respectively.

\begin{figure}[t]
\centering

\begin{subfigure}{0.48\linewidth}
    \centering
    \includegraphics[width=\linewidth]{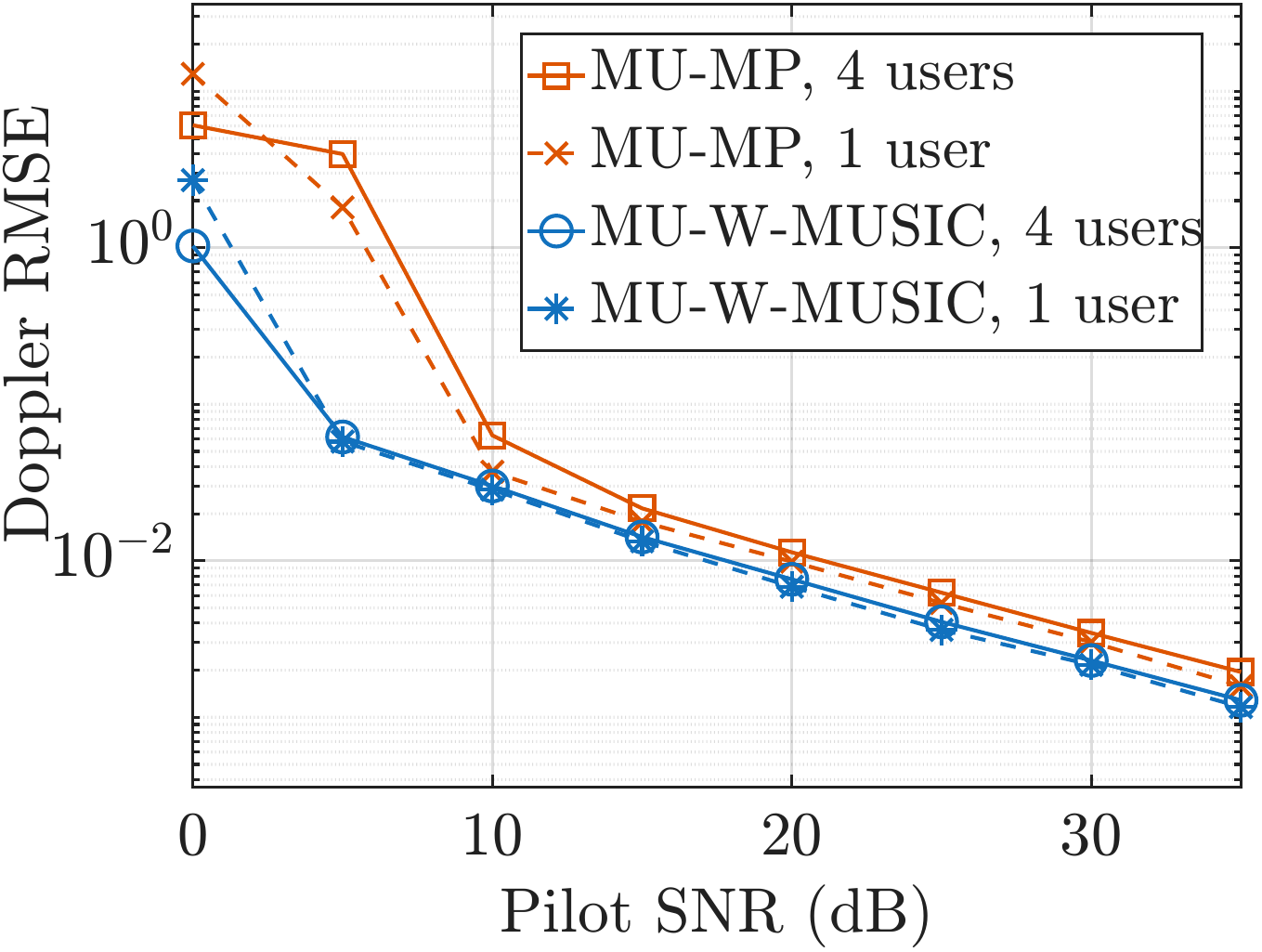}
    \caption{Doppler RMSE}
    \label{fig:rmse_Doppler}
\end{subfigure}
\hfill
\begin{subfigure}{0.48\linewidth}
    \centering
    \includegraphics[width=\linewidth]{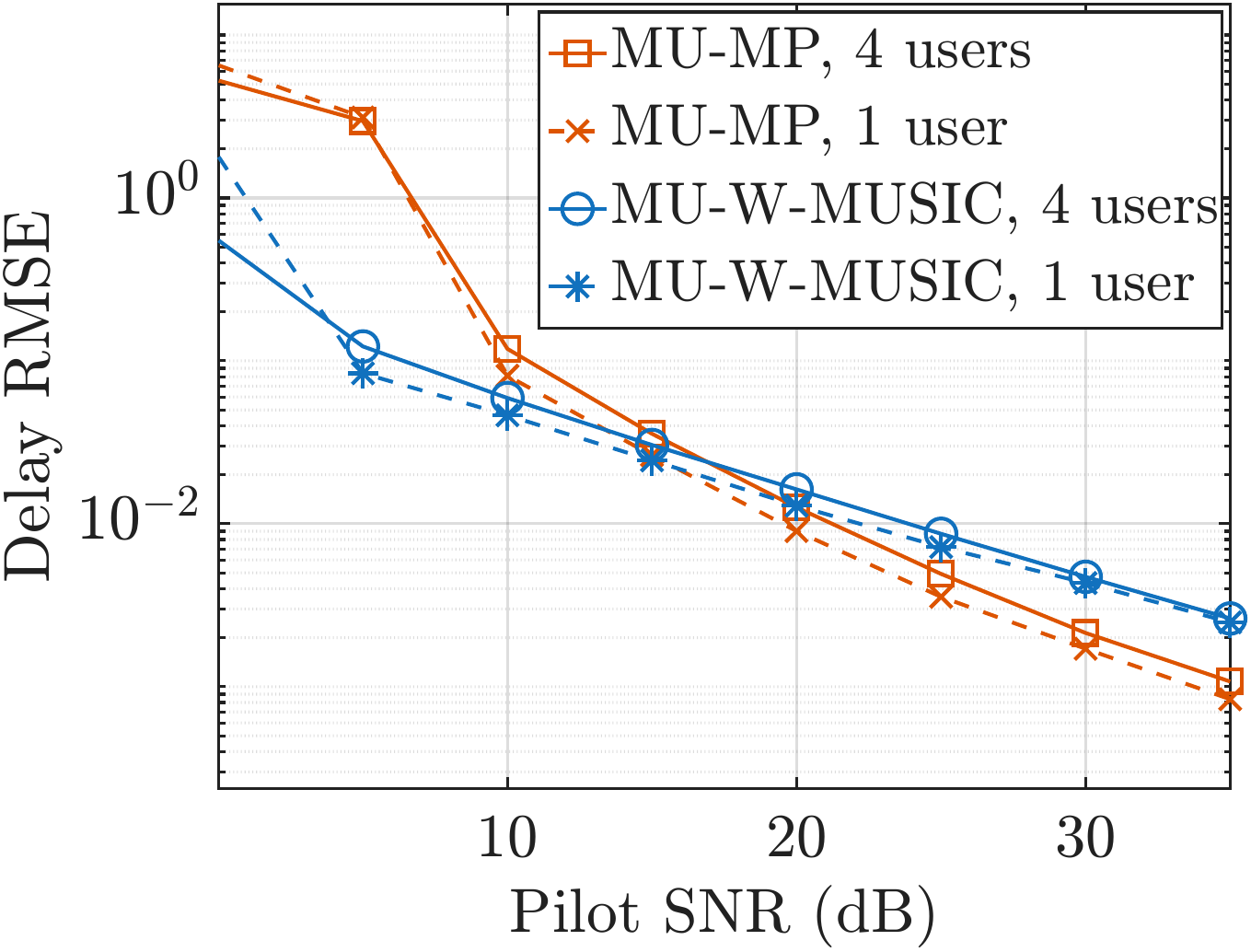}
    \caption{Delay RMSE}
    \label{fig:rmse_delay}
\end{subfigure}

\vspace{2mm}

\begin{subfigure}{0.48\linewidth}
    \centering
    \includegraphics[width=\linewidth]{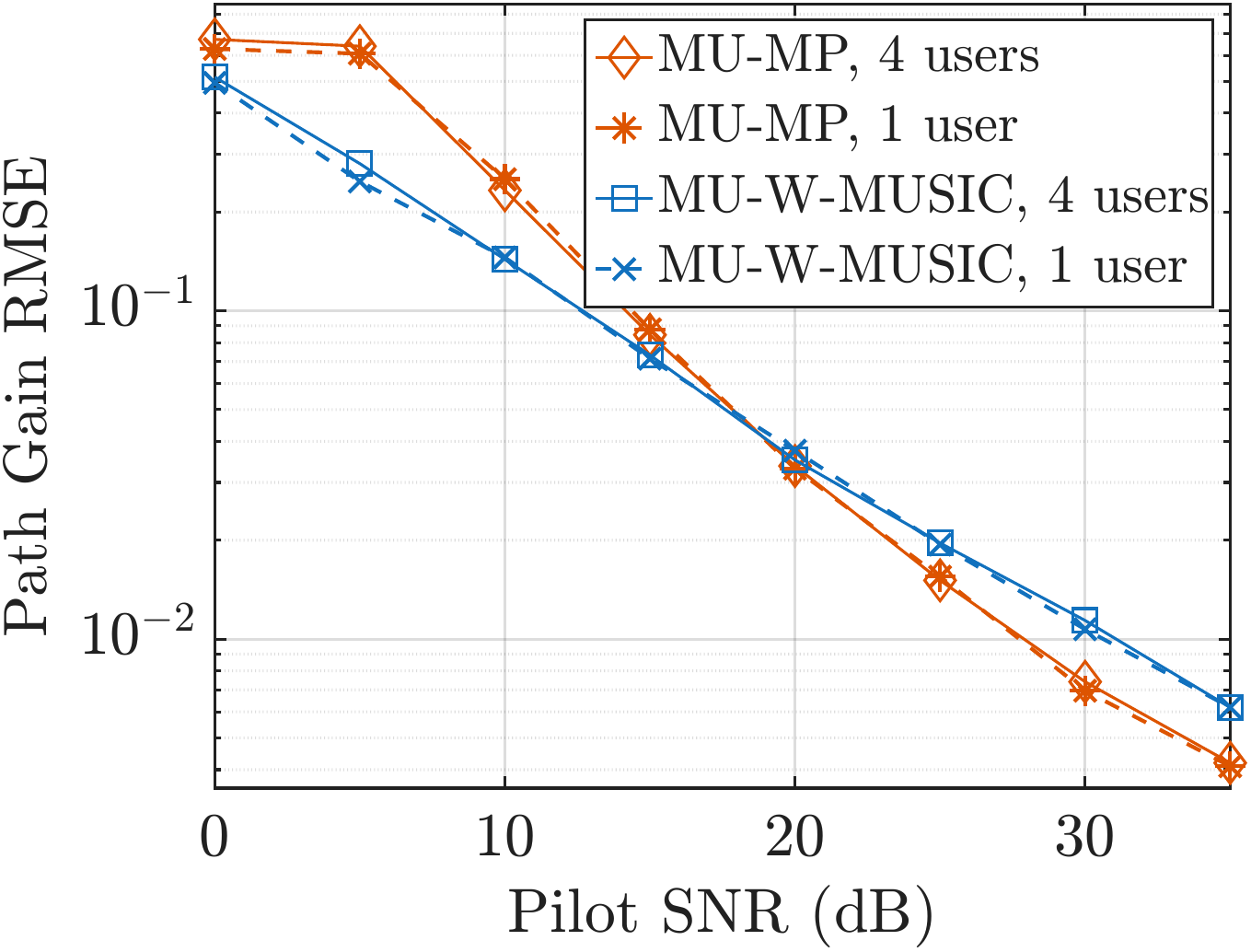}
    \caption{Path gain RMSE}
    \label{fig:rmse_pathgain}
\end{subfigure}
\hfill
\begin{subfigure}{0.48\linewidth}
    \centering
    \includegraphics[width=\linewidth]{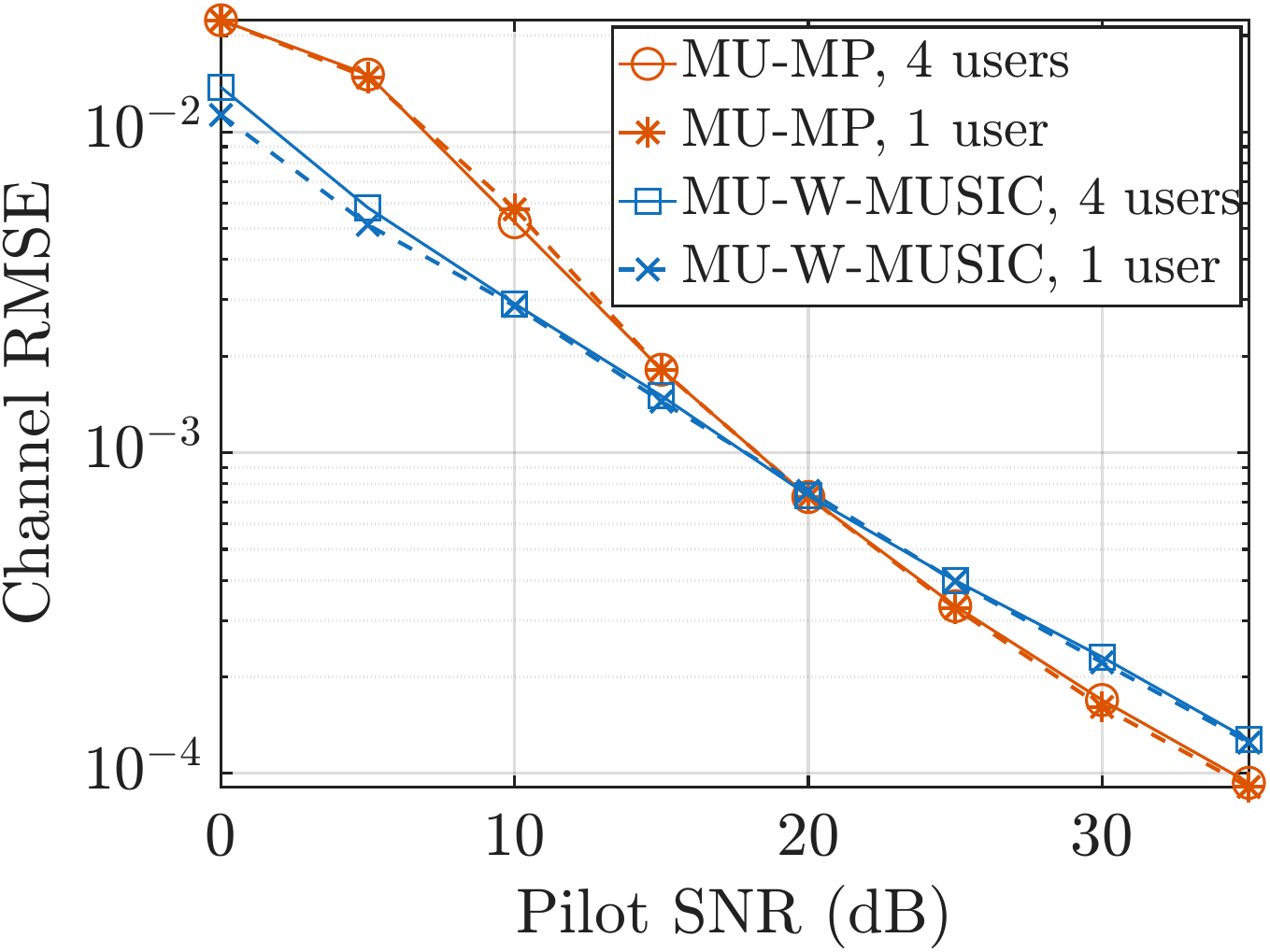}
    \caption{Channel RMSE}
    \label{fig:rmse_fullchannel}
\end{subfigure}

\caption{RMSE versus pilot SNR for Doppler, delay, path gain, and full channel estimation. 
\vspace{-0.5cm}
}
\label{fig:rmse_combined}

\end{figure}

Fig.~\ref{fig:rmse_Doppler} and Fig.~\ref{fig:rmse_delay} show the Doppler and delay RMSE against pilot SNR. In the low-SNR regime (0-5 dB), the proposed MU-W-MUSIC-based scheme outperforms the proposed MU-MP-based method in both Doppler and delay estimation. As the SNR increases beyond 10 dB, the Doppler performance gap narrows significantly, and both methods achieve comparable accuracy in the $10^{-3}$-$10^{-2}$ range at high SNR.

For delay estimation, a similar low-SNR trend is observed in Fig.~\ref{fig:rmse_delay}, where MU-W-MUSIC outperforms MU-MP. However, beyond 15 dB , MU-MP becomes more accurate. In the high-SNR region (25–35 dB), it consistently achieves lower delay RMSE. At an RMSE level of $10^{-3}$, MU-MP provides about a 5 dB SNR gain compared to MU-W-MUSIC.

Fig.~\ref{fig:rmse_pathgain} shows the path gain RMSE. Both methods exhibit comparable performance across the full SNR range. At low SNR (0–5 dB), MU-W-MUSIC has a slight advantage. Beyond 15 dB, the two methods overlap closely, and at high SNR (30-35 dB), both achieve RMSE on the order of below $10^{-2}$.

Fig.~\ref{fig:rmse_fullchannel} demonstrates the RMSE of the reconstructed full channel matrix, where the matrix form $\mathbf{H}$ follows the parametric model in \ref{sec:delay_doppler_channel_model}. The results show that both methods achieve full channel RMSE below $10^{-3}$ when the SNR exceeds 20 dB. MU-MP  achieves approximately 3 dB gain at the RMSE level around $10^{-4}$ compared to MP-W-MUSIC. 

In Fig.~\ref{fig:rmse_combined}, a slight performance loss is observed in delay and Doppler parameter estimation under the multi-user setting in Fig.~\ref{fig:rmse_Doppler} and Fig.~\ref{fig:rmse_delay}. For the single-user case, the proposed MU-W-MUSIC reduces to the W-MUSIC method in \cite{weighted_2d_music_delay_doppler}, and thus shares the same estimation structure and performance. However, the full channel reconstruction results remain almost identical between single-user and multi-user cases in Fig~\ref{fig:rmse_fullchannel}. 

Table~\ref{tab:complexity_comparison} quantifies the CMs of both methods. Under our simulation setup, MU-W-MUSIC requires $2.65 \times 10^{8}$ CMs, compared to $2.68 \times 10^{7}$ CMs for MU-MP, corresponding to approximately one order of magnitude reduction. This gap is dominated by the spectral sampling stage in MU-W-MUSIC, which is absent in MU-MP. In contrast, the main cost of MU-MP arises from the SVD. The overrall results confirms that our proposed MU-MP method achieves significantly lower complexity while maintaining comparable or improved estimation accuracy at moderate-to-high SNR.

\section{Conclusion} \label{sec:conclusion}
{ In this paper, we developed a multiuser OTFS channel parameter estimation framework based on MU-PCP. By multiplexing users along the Doppler dimension while preserving a separable exponential form, MU-PCP enables high-resolution estimation of fractional delay and Doppler. We  proposed novel approaches that exploit the observed Doppler and MU-PCP Doppler partitions to recover channel parameters, followed by channel gain reconstruction via a multiuser parametric dictionary. Building on this formulation, we extended W-MUSIC and MP to multiuser OTFS, resulting in two high-resolution approaches, with MP achieving a fully grid-independent implementation. Simulation results demonstrate accurate estimation of fractional delay, Doppler, and channel gains, enabling reliable reconstruction of the full channel matrix. Future work will investigate adaptive or hybrid strategies that combine the robustness of subspace-based methods with the high-resolution capabilities of parametric approaches.}
\begin{table}[t]
\centering
\caption{Complexity Comparison}
\label{tab:complexity_comparison}
\begin{tabular}{lcc}
\hline
\textbf{Operation} 
& \textbf{MU-W-MUSIC} 
& \textbf{MU-MP} \\
\hline

$C_\mathrm{cov}$ / $C_\mathrm{Hankel}$ 
& $4.61 \times 10^{6}$ 
& $7.06 \times 10^{4}$ \\

$C_\mathrm{EVD}$ / $C_\mathrm{SVD}$
& $3.28 \times 10^{7}$ 
& $2.59 \times 10^{7}$ \\

$C_{\mathrm{spec}}$ / --- 
& $2.10\times 10^{8}$ 
& --- \\

$C_\mathrm{WLS}$ / $C_{\mathbf{T}}$
& $1.09 \times 10^{6}$ 
& $8.50 \times 10^{5}$ \\

$C_{\mathrm{root}}$ / $C_\mathrm{EVD}$
& $1.33 \times 10^{5}$ 
& $1.73 \times 10^{3}$ \\

$C_{\mathrm{delay}}$ / $C_{\mathrm{proj}}$
& $1.97 \times 10^{7}$ 
& $2.46 \times 10^{4}$ \\
\hline

\textbf{Total CMs} 
& $\mathbf{2.65 \times 10^{8}}$ 
& $\mathbf{2.68 \times 10^{7}}$ \\
\hline
\end{tabular}
\vspace{-0.6cm}
\end{table}

\bibliographystyle{IEEEtran}
\bibliography{biblio}

\end{document}